\documentclass[%
reprint,
superscriptaddress,
longbibliography,
amsmath,
amssymb,
aps,
pra
]{revtex4-2}
\usepackage{float}
\usepackage{subfig}
\usepackage{mathrsfs}
\usepackage{pifont}
\usepackage{color}
\usepackage{microtype}
\usepackage{graphicx}
\usepackage{dcolumn}
\usepackage{bm}
\usepackage[mathlines]{lineno}
\usepackage{siunitx}
\usepackage[colorlinks=true,citecolor=blue,urlcolor=blue,linkcolor=blue]{hyperref}
\usepackage[T1]{fontenc}
\usepackage[utf8]{inputenc}
\usepackage[english]{babel}   
\usepackage{hyphenat}    
\allowdisplaybreaks

\begin{document}

\title{Theoretical study of high-order harmonic generation in solids based on deep-learning methods}

\author{Wen-Feng Huang}
\affiliation{\shortstack{Wuhan Institute of Physics and Mathematics, Innovation Academy for Precision}{ Measurement Science and Technology, Chinese Academy of Sciences, Wuhan 430071, China}}
\affiliation{\shortstack{School of Physical Sciences, University of Chinese Academy of Sciences, Beijing 100049, China}}

\author{Jia-Qi Liu}
\affiliation{\shortstack{Wuhan Institute of Physics and Mathematics, Innovation Academy for Precision}{ Measurement Science and Technology, Chinese Academy of Sciences, Wuhan 430071, China}}

\author{Xue-Bin Bian}
\email{Contact author: xuebin.bian@wipm.ac.cn}
\affiliation{\shortstack{Wuhan Institute of Physics and Mathematics, Innovation Academy for Precision}{ Measurement Science and Technology, Chinese Academy of Sciences, Wuhan 430071, China}}

\begin{abstract}
High-order harmonic generation (HHG) in solids provides a compact and coherent EUV radiation, which has attracted considerable attention. Many theoretical methods for laser-crystal interactions are developed. However, an approach to treat large-scale amorphous systems with \textit{ab-initio} accuracy is still lacking because of the enormous computational cost. Here, we incorporate molecular dynamics and tight-binding electron Hamiltonians based on deep-learning methods to solve the problem efficiently. Its accuracy is validated by comparing the HHG spectra of monolayer hexagonal boron nitride (h-BN) with those obtained using existing methods.
Then we use it to study the temperature-dependent HHG in bulk Si. We uncover a distinct harmonic-order-dependent thermal response: low-order harmonics are enhanced with increasing temperature, whereas high-order harmonics are suppressed. The former is caused by the modification of the inter-site hopping in the system, while the latter is mainly influenced by the structural disorder induced by the thermal effects, in agreement with experimental observations. This work establishes a scalable computational strategy for investigating HHG in solids and provides new insights into the interplay between lattice fluctuations and ultrafast electron dynamics.
\begin{hyphenrules}{nohyphenation}

\end{hyphenrules}

\end{abstract}
\maketitle
\section{Introduction}\label{section1}
Strong-field light-matter interactions enable the generation and manipulation of non-equilibrium electronic dynamics on the intrinsic electronic timescale. High-order harmonic generation (HHG), as a highly nonlinear coherent radiation process, has emerged as an indispensable tool for probing electronic motion. In gaseous systems, HHG has been successfully applied to areas such as attosecond pulse generation and molecular orbital reconstruction \cite{RN1,RN2,RN3,RN4,RN5,RN6,RN7,RN8,RN9,RN10,RN11,RN12}. Compared with gaseous media, solids exhibit much higher particle densities and, in crystalline materials, well-defined electronic band structures \cite{RN13,RN14,RN15,RN17,RN18}. These characteristics enable distinct strong-field electron dynamics, making solid-state HHG a powerful probe of ultrafast electronic processes in condensed matter. 

In recent years, experimental and theoretical investigations of solid-state HHG have advanced rapidly. Experimentally, measurements of HHG in semiconductor materials such as ZnO have disclosed intrinsic correlations between harmonic spectral features and underlying electronic band structures \cite{RN14}, demonstrating that HHG can serve as a vital spectroscopic tool for probing electronic structures and ultrafast dynamics in solids. However, most existing studies on solid-state HHG have focused on ideal crystal structures \cite{rn21,AA1,rn23} or fixed temperature conditions \cite{rn24}. In realistic materials, lattice thermal motion is inevitable at finite temperatures. Previous theoretical studies have demonstrated that lattice thermal fluctuations can modify the local electronic environment, potentially affecting electronic transitions \cite{rn25,rn26}. Recent experiments by Mokhtari \textit{et al.} \cite{2604.07543} have further revealed temperature-dependent variations of harmonic intensities in bulk Si, where certain harmonic orders decrease with increasing temperature. These findings highlight the importance of developing computational approaches capable of describing finite-temperature HHG responses under realistic material conditions.

To accurately describe finite-temperature effects, efficient sampling of thermally fluctuating atomic configurations is crucial. In recent years, machine-learning methods have been introduced to study the strong-field physics \cite{AA2,AA3}. Deep-learning potential-based molecular dynamics (DPMD) has provided an effective approach for large-scale atomistic simulations by integrating the high accuracy of first-principles calculations with the computational efficiency of classical molecular dynamics \cite{BB1,BB2}. Among various machine-learning interatomic potentials, neuroevolution potentials (NEP) implemented in the GPUMD package exhibit excellent accuracy and computational efficiency \cite{10.1063/5.0106617,rn30,rn31}, enabling efficient sampling of temperature-dependent atomic configurations in complex material systems.

At present, finite-temperature investigations of solid-state HHG remain heavily constrained by computational methodologies. One-dimensional models, while capable of illustrating specific fundamental physics, fail to capture the complex electronic structures inherent to real three-dimensional materials \cite{rn25}. Time-dependent density functional theory (TDDFT) provides an accurate description of electron dynamics. However, its computational cost becomes prohibitive when applied to systems containing large numbers of atoms or during extensive thermal configuration sampling \cite{rn26}. On the other hand, conventional tight-binding models offer high computational efficiency but typically rely on empirical parameters \cite{rn27,rn28}, rendering them ill-suited for describing complex structural perturbations induced by temperature variations.

Recently, deep-learning tight-binding methods, such as DeePTB \cite{rn29,RU101}, have emerged as a compelling approach that offers a new avenue for addressing these challenges. By training on high-precision first-principles density functional theory (DFT) calculations, DeePTB constructs parametrized tight-binding Hamiltonians that rapidly predict electronic structures across diverse structural configurations, effectively bridging the gap between first-principles accuracy and large-scale computational efficiency. Furthermore, when constructed using an orthogonal localized orbital basis, the resulting deep-learning tight-binding Hamiltonians exhibit sparsity and favorable scalability, while providing a compact effective description of the relevant electronic degrees of freedom \cite{Zou_2026}. Combining DeePTB with the semiconductor Bloch equations (SBE) enables the simulation of electron dynamics driven by strong laser fields at finite temperatures, providing a promising route to systematically elucidate temperature-dependent solid-state HHG mechanisms.

In this work, the theoretical methods are presented in Sec. \ref{section2}, and the results are shown in Sec. \ref{section3}. We summarize the paper in Sec. \ref{section4}.
\section{Theory}\label{section2}
In this section, we introduce the methods used in our calculations. Subsection \ref{subsection2A} describes the Hamiltonian matrix obtained by DeePTB, the others illustrate the time evolution of the system when the laser field is added.
\subsection{Deep-Learning Tight-Binding Models}\label{subsection2A}
A distinct advantage of the deep-learning tight-binding model implemented in DeePTB, compared to other deep-learning electronic structure prediction approaches, is that it directly yields the Hamiltonian matrix in the Wannier gauge \cite{rn29}. The Wannier basis set offers a significantly lower dimensionality than other linear combinations of atomic orbitals basis sets, while possessing orthogonal and sparse Hamiltonian properties \cite{MOSTOFI2008685}. These features create favorable conditions for accelerating subsequent time-evolution calculations.

For a supercell containing \(N\) primitive cells, the Bravais lattice vectors can be expressed as \(\boldsymbol{R}_n = \sum_{i=1}^{3} n_i \boldsymbol{a}_i\), where \(\boldsymbol{a}_i\) represents the primitive lattice vectors. A set of localized Wannier functions is given by \(|\boldsymbol{R} m\rangle\), where \(m\) indexes the Wannier orbital. The set of Wannier functions forms an orthonormal basis, satisfying \(\langle \boldsymbol{R}n | \boldsymbol0m \rangle = \delta_{\textbf{0}\boldsymbol{R}} \delta_{mn}\). Then the Bloch-like functions are defined as
\begin{equation}
\label{eq:bloch}
|\phi_{m\boldsymbol{k}}^{(W)}\rangle = \frac{1}{\sqrt{N}} \sum_{\boldsymbol{R}} e^{i \boldsymbol{k} \cdot \boldsymbol{R}} |\boldsymbol{R}m\rangle,
\end{equation}
and the Hamiltonian and dipole in the Wannier gauge are given by \cite{rn28,rn34}
\begin{equation}
\label{eq:hamiltonian}
H_{mn,\boldsymbol{k}}^{(W)}
=
\langle \phi_{m\boldsymbol{k}}^{(W)}|\hat{H}_0|\phi_{n\boldsymbol{k}}^{(W)}\rangle
\equiv
\sum_{\boldsymbol{R}}e^{i\boldsymbol{k}\cdot\boldsymbol{R}}
\langle \boldsymbol{0}m|\hat{H}_0|{\boldsymbol{R}n}\rangle,
\end{equation}
\begin{equation}
\label{eq:dipole}
\boldsymbol{D}_{mn,\boldsymbol{k}}^{(W)}
=
\langle u_{m\boldsymbol{k}}^{(W)}
|i\nabla_{\boldsymbol{k}}|
u_{n\boldsymbol{k}}^{(W)}\rangle
\equiv
\sum_{\boldsymbol{R}} e^{i\boldsymbol{k}\cdot\boldsymbol{R}}
\langle \boldsymbol{0}m|\hat{\boldsymbol{r}}|{\boldsymbol{R}n}\rangle,
\end{equation}
where \(\boldsymbol{R}\) runs over all lattice vectors, \(|u_{n\boldsymbol{k}}^{(W)}\rangle\) is the periodic part of the Bloch-like functions, \(\hat{H}_0\) and \(\boldsymbol{\hat{r}}\) are the unperturbed Hamiltonian and position operator. The superscript \((\text{W})\) specifies the Wannier gauge. During the band-fitting and prediction processes in DeePTB, the centers of the Wannier orbitals do not shift relative to the atomic positions. We can transform the system into the Hamiltonian gauge by diagonalizing \(H^{({W})}(\boldsymbol{k})\) as \cite{rn28,rn34}

\begin{equation}
\label{eq:diagonalization}
{H}^{(H)}(\boldsymbol{k}) = U^\dagger(\boldsymbol{k}) H^{({W})}(\boldsymbol{k}) U(\boldsymbol{k}) ,
\end{equation}
and the corresponding matrix \(\boldsymbol{D}^{(H)}(k)\) can be given by
\begin{equation}
\label{eq:diagonalization_dipole}
\boldsymbol{D}^{({H})}(\boldsymbol{k}) = U^\dagger(\boldsymbol{k}) \boldsymbol{D}^{(W)}(\boldsymbol{k}) U(\boldsymbol{k})
+
iU^\dagger(\boldsymbol{k})\frac{\partial}{\partial\boldsymbol{k}}U(\boldsymbol{k}),
\end{equation}
where \(H^{(H)}\) and \(\boldsymbol{D}^{(H)}\) are the Hamiltonian matrix and transition dipole matrix under the Bloch basis, and \(U(k)\) is the unitary matrix that diagonalizes \(H^{(W)}\). The corresponding Bloch states in the Hamiltonian gauge are given by 
\begin{equation}
|\phi^{(H)}_{m\boldsymbol{k}}\rangle = \sum_{m} U_{mn}(\boldsymbol{k}) |\phi^{(W)}_{n\boldsymbol{k}}\rangle.
\end{equation}

In this work, we employ the DeePTB approach to construct the effective electronic Hamiltonian of the system, whose matrix element is \(\langle \boldsymbol{0}m|\hat{H}_0|{\boldsymbol{R}n}\rangle\). Based on the Slater-Koster formalism, DeePTB combines conventional tight-binding theory with neural networks by introducing environment-dependent corrections to the tight-binding parameters. This formulation goes beyond the conventional two-center approximation, enabling a more accurate description of the influence of local structural variations on electronic interactions \cite{Zou_2026}. Furthermore, DeePTB employs local atomic environment descriptors that are invariant under translations, rotations, and permutations of equivalent atoms \cite{jmi.2025.17}, while the Slater-Koster formalism ensures the proper transformation of the Hamiltonian under spatial rotations \cite{FF01, Zou_2026}. The tight-binding parameters are optimized using electronic eigenvalues obtained from first-principles calculations, allowing the resulting Hamiltonian to capture the electronic structure in complex local atomic environments while preserving the underlying physical symmetries and orbital-coupling structure.

\subsection{Density-Matrix-Based SBE}\label{subsection2B}
To describe the interaction between an intense laser field and an ideal crystal, we start with the single-particle time-dependent Schrödinger equation (TDSE). Within the electric dipole approximation and velocity gauge \cite{rn32}, the time evolution of electrons driven by the laser field is governed by 

\begin{equation}
\begin{aligned}
i \frac{\partial \psi(\boldsymbol{r}, t)}{\partial t} &= \hat{H}(t)\psi(\boldsymbol{r},t)\\&= \left[ \frac{1}{2} \left( \boldsymbol{p} + \textbf{A}(t) \right)^2 + V(\boldsymbol{r}) \right] \psi(\boldsymbol{r}, t)\\
&=[\hat{H}_0+\boldsymbol{p}\cdot\textbf{A}(t)+\frac{1}{2}\textbf{A}^2(t)]\psi(\boldsymbol{r}, t),
\end{aligned}
\label{eq:schrodinger_velocity}
\end{equation}
where \(\textbf{A}(t)\) is the vector potential of the laser field, \(V(\boldsymbol{r})\) represents the periodic potential of the crystal. Using the acceleration theorem for Bloch electrons \cite{rn35,rn36,rn37}, we construct an adiabatic instantaneous eigenstate of Eq. (\ref{eq:schrodinger_velocity}). The Houston states can be defined as 

\begin{equation}
\label{eq:houston_states}
\tilde\phi_{n\boldsymbol{k}}(\boldsymbol{r}) = e^{-i \textbf{A}(t) \cdot \boldsymbol{r}} \phi_{n\boldsymbol{k}}(\boldsymbol{r}),
\end{equation}
which satisfy the time-independent eigenvalue equation:

\begin{equation}
\label{eq:houston_eigenvalue}
\hat{H}(t)\tilde{\phi}_{n\boldsymbol{k}}(\boldsymbol{r},t)
=
\varepsilon_{n\boldsymbol{k}}\tilde{\phi}_{n\boldsymbol{k}}(\boldsymbol{r},t).
\end{equation}
Then, the electronic wave function can be expanded in terms of these states:
\begin{equation}
\begin{aligned}
\psi{(\boldsymbol{r},t)}
&=\int_{BZ}\sum_n{C_{n\boldsymbol{k}}(t)\tilde{\phi}_{n\boldsymbol{k}}}(\boldsymbol{r},t)d\boldsymbol{k}
\\&= \int_{BZ}\sum_n{C_{n\boldsymbol{k}}(t)e^{-i\boldsymbol{A}\cdot\boldsymbol{r}}\phi_{n\boldsymbol{k}}}(\boldsymbol{r})d\boldsymbol{k}.
\end{aligned}
\label{eq:wavefunction_expansion}
\end{equation}
The corresponding density matrix elements are defined as

\begin{equation}
\label{eq:density_matrix_def}
\rho_{nm}^{\boldsymbol{k}}(t)= C^{*\boldsymbol{k}}_m(t) C^{\boldsymbol{k}}_n(t).
\end{equation}
Substituting Eqs. (\ref{eq:houston_states})--(\ref{eq:density_matrix_def}) into the TDSE yields the SBE in the Hamiltonian gauge:
\begin{equation}
\small
\begin{aligned}
\frac{\partial \rho_{nm}^{(H)\boldsymbol{k}_0}(t)}{\partial t} &= i[\rho^{(H)\boldsymbol{k}_0}_{nl}(t) H^{(H)\boldsymbol{k}(t)}_{lm}-H^{(H)\boldsymbol{k}(t)}_{nl}\rho^{(H)\boldsymbol{k}_0}_{lm}(t)]\\&
+i\textbf{E}(t)\cdot[\rho^{(H)\boldsymbol{k}_0}_{nl}(t){\boldsymbol{D}^{(H)\boldsymbol{k}(t)}_{lm}}-{\boldsymbol{D}^{(H)\boldsymbol{k}(t)}_{nl}\rho^{(H)\boldsymbol{k}_0}_{lm}(t)}].
\end{aligned}
\label{eq:sbe_wannier}
\end{equation}
Then we can get the SBE in Wannier gauge through unitary transformation:
\begin{equation}
\small
\begin{aligned}
\frac{\partial \rho_{nm}^{(W)\boldsymbol{k}_0}(t)}{\partial t} &= i\sum_{l}[\rho^{(W)\boldsymbol{k}_0}_{nl} (t)H^{(W)\boldsymbol{k}(t)}_{lm}-H^{(W)\boldsymbol{k}(t)}_{nl}\rho^{(W)\boldsymbol{k}_0}_{lm}(t)]\\&+i\textbf{E}(t)\cdot[\rho^{(W)\boldsymbol{k}_0}_{nl}(t){\boldsymbol{D}^{(W)\boldsymbol{k}(t)}_{lm}}-{\boldsymbol{D}^{(W)\boldsymbol{k}(t)}_{nl}}\rho^{(W)\boldsymbol{k}_0}_{lm}(t)],
\end{aligned}
\label{eq:sbe_wannier}
\end{equation}
where \(\textbf{E}(t) = -\frac{\partial \textbf{A}(t)}{\partial t}\) is the electric field of the laser pulse, the crystal momentum evolves as \(\boldsymbol{k}(t) = \boldsymbol{k}_0 + \textbf{A}(t)\), and \(\boldsymbol{k}_0\) denotes the initial crystal momentum \cite{RN1991,RN1992,RN1993}.

In SBE, phenomenological dephasing terms are typically introduced to account for dephasing effect induced by electron--electron and electron--phonon scattering, thereby yielding theoretical results in better agreement with experimental observations \cite{RN15}. Numerically, the dephasing relaxation is handled naturally within the Hamiltonian gauge. Specifically, we propagate the coherent evolution equations in the Wannier gauge \cite{rn28} while incorporating the dephasing terms in the Hamiltonian gauge. As shown in Eq. (\ref{eq:sbe_hamiltonian}), the equations of motion in the Hamiltonian gauge are formulated as \cite{rn27}

\begin{equation}
\small
\begin{aligned}
\frac{\partial \rho_{nm}^{(H)\boldsymbol{k}_0}(t)}{\partial t} &= i[\rho^{(H)\boldsymbol{k}_0}_{nl} (t)H^{(H)\boldsymbol{k}(t)}_{lm}-H^{(H)\boldsymbol{k}(t)}_{nl}\rho^{(H)\boldsymbol{k}_0}_{lm}(t)]\\&
+i\textbf{E}(t)\cdot[\rho^{(H)\boldsymbol{k}_0}_{nl}(t){\boldsymbol{D}^{(H)\boldsymbol{k}(t)}_{lm}}-{\boldsymbol{D}^{(H)\boldsymbol{k}(t)}_{nl}\rho^{(H)\boldsymbol{k}_0}_{lm}(t)}]\\&
-\frac{1-\delta_{nm}}{T_2}\rho^{(H)\boldsymbol{k}_0}_{nm}(t).
\end{aligned}
\label{eq:sbe_hamiltonian}
\end{equation}
The Crank-Nicolson scheme is used for the time evolution, and the total current is given by 

\begin{equation}
\begin{aligned}
\boldsymbol{J}(t) &= - \int_{BZ}\text{Tr}[\rho^{(W)\boldsymbol{k}_0}(t)\boldsymbol{v}^{(W)\boldsymbol{k}_0}(t)]d\boldsymbol{k}_0,
\end{aligned}
\label{eq:total_current}
\end{equation}
where the elements of the velocity matrix \(\boldsymbol{v}^{(W)\boldsymbol{k}_0}(t)\) which can be calculated from Heisenberg equation of motion is decomposed as 

\begin{equation}
\label{eq:velocity_elements}
\begin{aligned}
\boldsymbol{v}^{(W)\boldsymbol{k}_0}(t) =\frac{\partial{H}^{(W)\boldsymbol{k}(t)}}{\partial\boldsymbol{k}(t)}-i[\boldsymbol{D}^{(W)\boldsymbol{k}(t)},H^{(W)\boldsymbol{k}(t)}].
\end{aligned}
\end{equation}

However, we find that the matrix elements of the position operator are often
neglected in previous works \cite{GG01,GG02}. In this case, \(\boldsymbol{D}^{(W)\boldsymbol{k}(t)}_{mn}\) in Eq. (\ref{eq:dipole}) is zero and Eq. (\ref{eq:sbe_wannier}) becomes
\begin{equation}
\begin{aligned}
\frac{\partial \rho_{nm}^{(W)\boldsymbol{k}_0}(t)}{\partial t} &= i\sum_{l}[\rho^{(W)\boldsymbol{k}_0}_{nl} (t)H^{(W)\boldsymbol{k}(t)}_{lm}-H^{(W)\boldsymbol{k}(t)}_{nl}\rho^{(W)\boldsymbol{k}_0}_{lm}(t)],
\end{aligned}
\label{eq:sbe_wannier_adjust}
\end{equation}
and the time-dependent current becomes
\begin{equation}
\begin{aligned}
\boldsymbol{J}(t) &= - \int_{BZ} \text{Tr}[\rho^{(W)\boldsymbol{k}_0}(t)\frac{\partial{H}^{(W)\boldsymbol{k}(t)}}{\partial\boldsymbol{k}(t)}]d\boldsymbol{k}_0.
\end{aligned}
\label{eq:total_current_adjust}
\end{equation}
The HHG intensity spectrum \(I(\omega)\) is proportional to the squared magnitude of the Fourier transform of the time derivative of the macroscopic current as

\begin{equation}
\label{eq:hhg_spectrum}
I(\omega) \propto \left| \mathcal{F} \left[ \boldsymbol{J}(t)\right] \right|^2.
\end{equation}
\subsection{SBE evolution for occupied states}\label{subsection2C}

To reduce the computational cost and memory requirements, the SBE can alternatively be solved by propagating the single-particle states originating from the initially occupied Bloch states, without explicitly evolving the full density matrix. From Eq. (\ref{eq:wavefunction_expansion}), the expansion coefficient vector in the Houston basis
is defined as
\begin{equation}
\mathscr{C}_{\lambda\bm{k}}^{(H)}(t)
=
\begin{pmatrix}
\label{eq.define_c}
C_1^{\bm{k}}(t)\\
C_2^{\bm{k}}(t)\\
\vdots\\
C_N^{\bm{k}}(t)
\end{pmatrix}.
\end{equation}
Here, $N$ denotes the total number of energy bands,
and $\lambda$ labels the electrons participating in the
time evolution. According to the Pauli exclusion
principle \cite{JJ01,JJ02} and the identity principle \cite{JJ03},
$\lambda$ also serves as the corresponding valence-band index.

By applying a unitary transformation, the expansion
coefficients can be expressed in the Houston basis
under the Wannier gauge as
\begin{equation}
\label{eq.transmit_c}
\mathscr{C}_{\lambda\bm{k}_0}^{(W)}(t)
=
U^{\dagger}(\bm{k}(t))
\mathscr{C}_{\lambda\bm{k_0}}^{(H)}(t).
\end{equation}
Substituting Eq. (\ref{eq:houston_states}), Eq. (\ref{eq.define_c}) and Eq. (\ref{eq.transmit_c}) into Eq. (\ref{eq:schrodinger_velocity}), we obtain
\begin{equation}
\label{eq.evolution_current}
i\frac{\partial
\mathscr{C}_{\lambda\bm{k}_0}^{(W)}(t)}
{\partial t}
=
\left[
H^{(W)\bm{k}(t)}
+
\bm{E}(t)\cdot
\bm{D}^{(W)\bm{k}(t)}
\right]
\mathscr{C}_{\lambda\bm{k}_0}^{(W)}(t).
\end{equation}
It is worth noting that Eq. (\ref{eq.evolution_current}) and Eq. (\ref{eq:sbe_wannier}) are essentially equivalent. The Crank-Nicolson scheme is used for the time evolution too. Consequently, the electric current can be
expressed as
\begin{equation}
\begin{aligned}
\bm{J}(t)
&=
-\int_{\mathrm{BZ}}
\sum_{\lambda\in N_{\mathrm{occ}}}
\mathscr{C}_{\lambda\bm{k}_0}^{(W)\dagger}(t)
\boldsymbol{v}^{(W)\boldsymbol{k}_0}(t)
\mathscr{C}_{\lambda\bm{k}_0}^{(W)}(t)
\,\mathrm{d}\bm{k}_0,
\end{aligned}
\label{eq:current_occupy}
\end{equation}
where \(N_{occ}\) denotes the number of initially occupied states. As discussed in subsection \ref{subsection2B}, we ignore the term \(\boldsymbol{D}^{(W)\boldsymbol{k}(t)}\). Then we can get
\begin{equation}
\begin{aligned}
\bm{J}(t)
&=
-\int_{\mathrm{BZ}}
\sum_{\lambda\in N_{\mathrm{occ}}}
\mathscr{C}_{\lambda\bm{k}_0}^{(W)\dagger}(t)
\frac{\partial{H}^{(W)\boldsymbol{k}(t)}}{\partial\boldsymbol{k}(t)}
\mathscr{C}_{\lambda\bm{k}_0}^{(W)}(t)
\,\mathrm{d}\bm{k}_0.
\end{aligned}
\label{eq:current_occupy_adjust}
\end{equation}
\section{RESULTS AND DISCUSSION}\label{section3}
\subsection{Validation of DeePTB model for solid HHG in hexagonal boron nitride (h-BN)}
Prior to performing large-scale simulations at finite temperatures, we first validate the reliability of DeePTB in constructing the electronic Hamiltonian to describe the strong-field response of solids. h-BN is chosen as a benchmark system due to the availability of extensive theoretical studies for comparing its strong-field response characteristics \cite{rn28,ru100}. We performed DFT calculations using ABACUS package with the pseudopotentials and orbitals from APNS to generate the electronic structure dataset of the h-BN system \cite{rn39,rn40,rn41}, which was subsequently used to train the DeePTB model. We used the PBE functional for ABACUS calculations \cite{DD01}. 

\begin{figure}[!tbp]
\includegraphics[width=\linewidth]{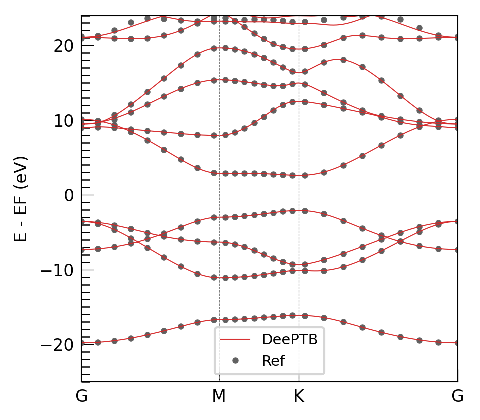}
\caption{\label{fig:hbn_band} Band structure of h-BN primitive cell given by DeePTB Hamiltonian (red lines). The black dots correspond to the band structure calculated by ABACUS. EF means the Fermi energy. }
\end{figure}

Figure \ref{fig:hbn_band}  compares the band structure of h-BN obtained by DeePTB with the DFT calculation results. The DeePTB model accurately reproduces the main band features of the system, with particularly good agreement in the region near the Fermi level. This indicates that DeePTB is capable of effectively learning the electronic coupling relationships in the h-BN system, providing a reliable Hamiltonian for subsequent strong-field electron dynamics calculations.

\begin{figure*}[!tbp]

\centering

\subfloat{
    \includegraphics[
    width=0.32\textwidth,
    trim=40 15 90 40,
    clip
    ]{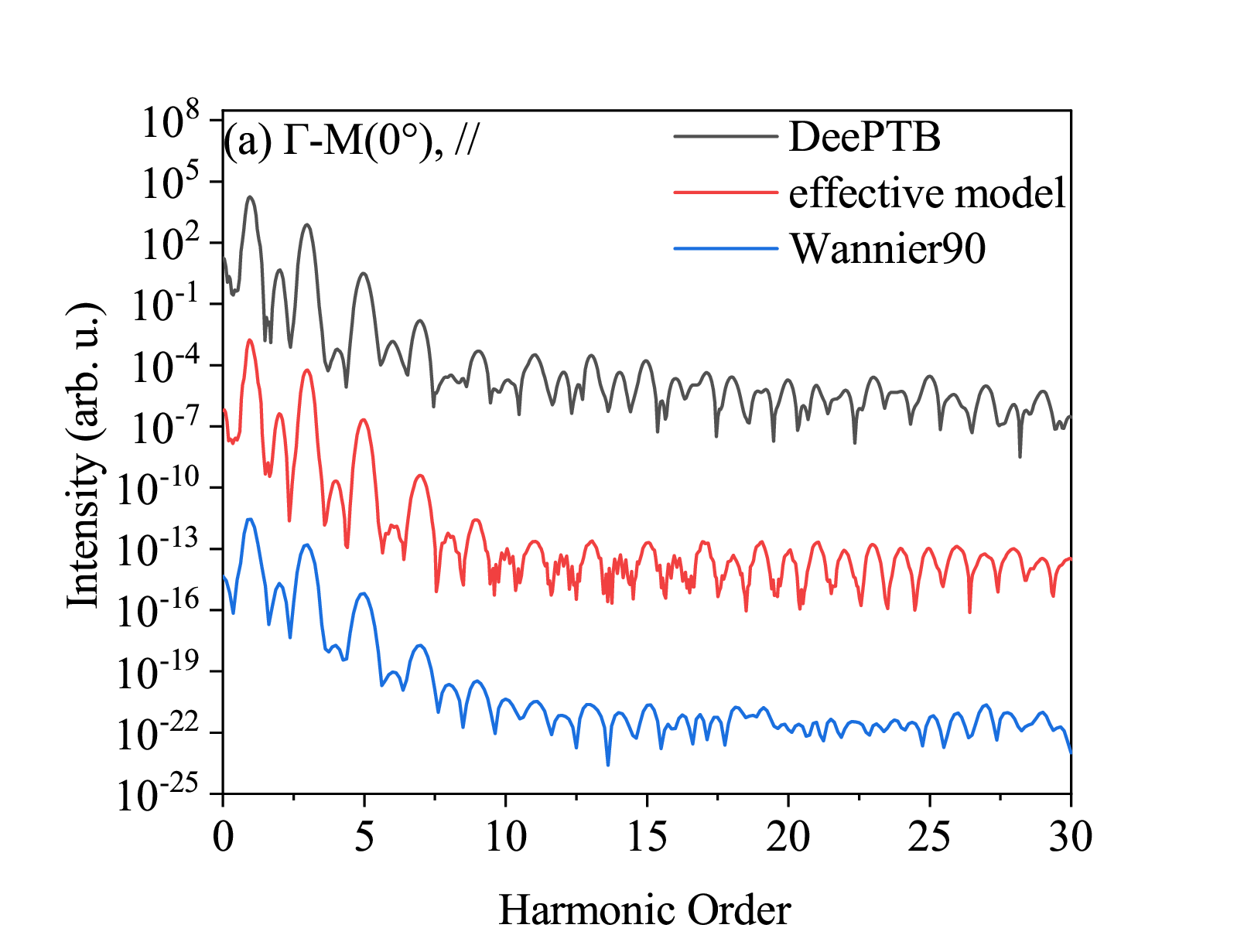}
}
\hfill
\subfloat{
    \includegraphics[
    width=0.32\textwidth,
    trim=40 15 90 40,
    clip
    ]{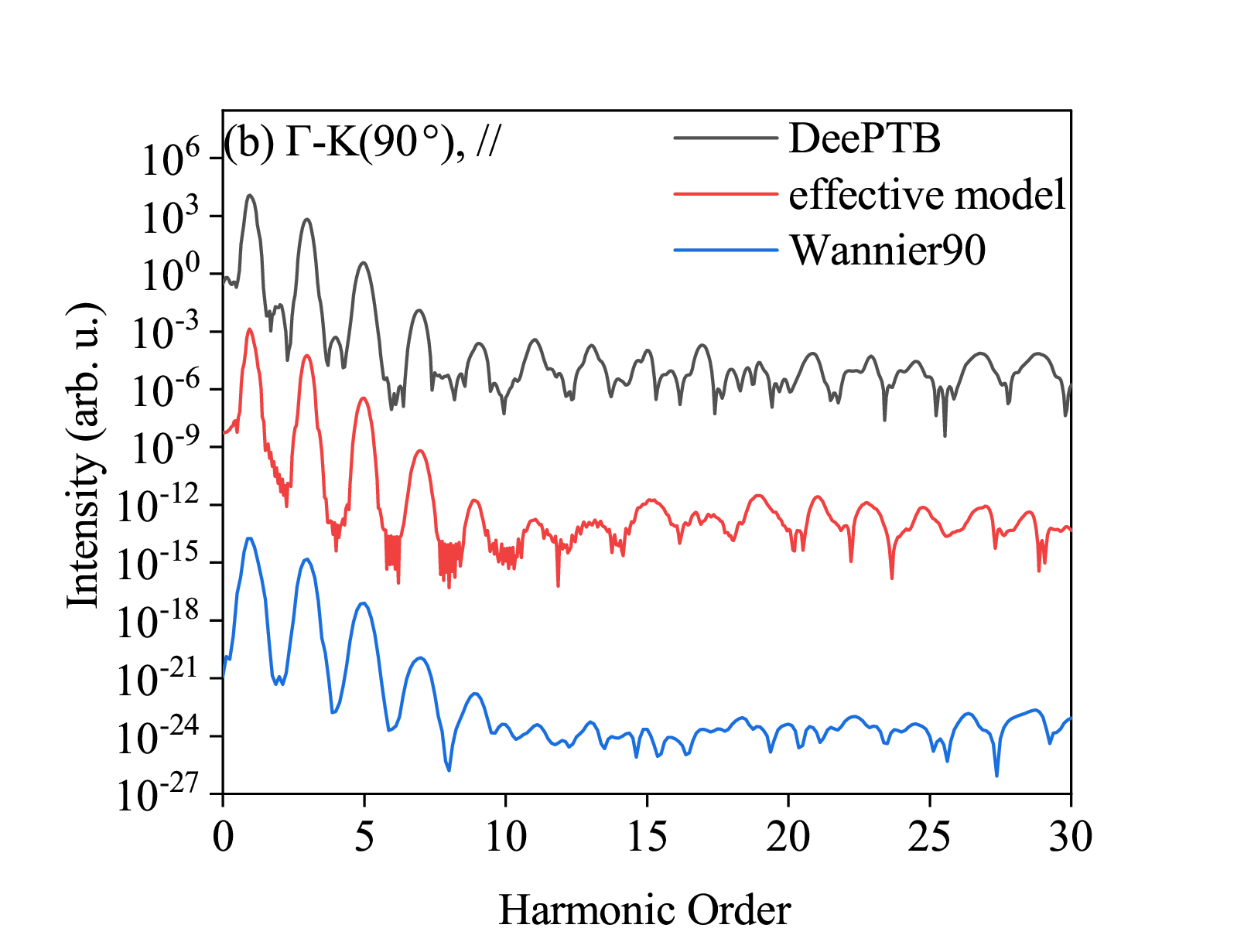}
}
\hfill
\subfloat{
    \includegraphics[
    width=0.32\textwidth,
    trim=40 15 90 40,
    clip
    ]{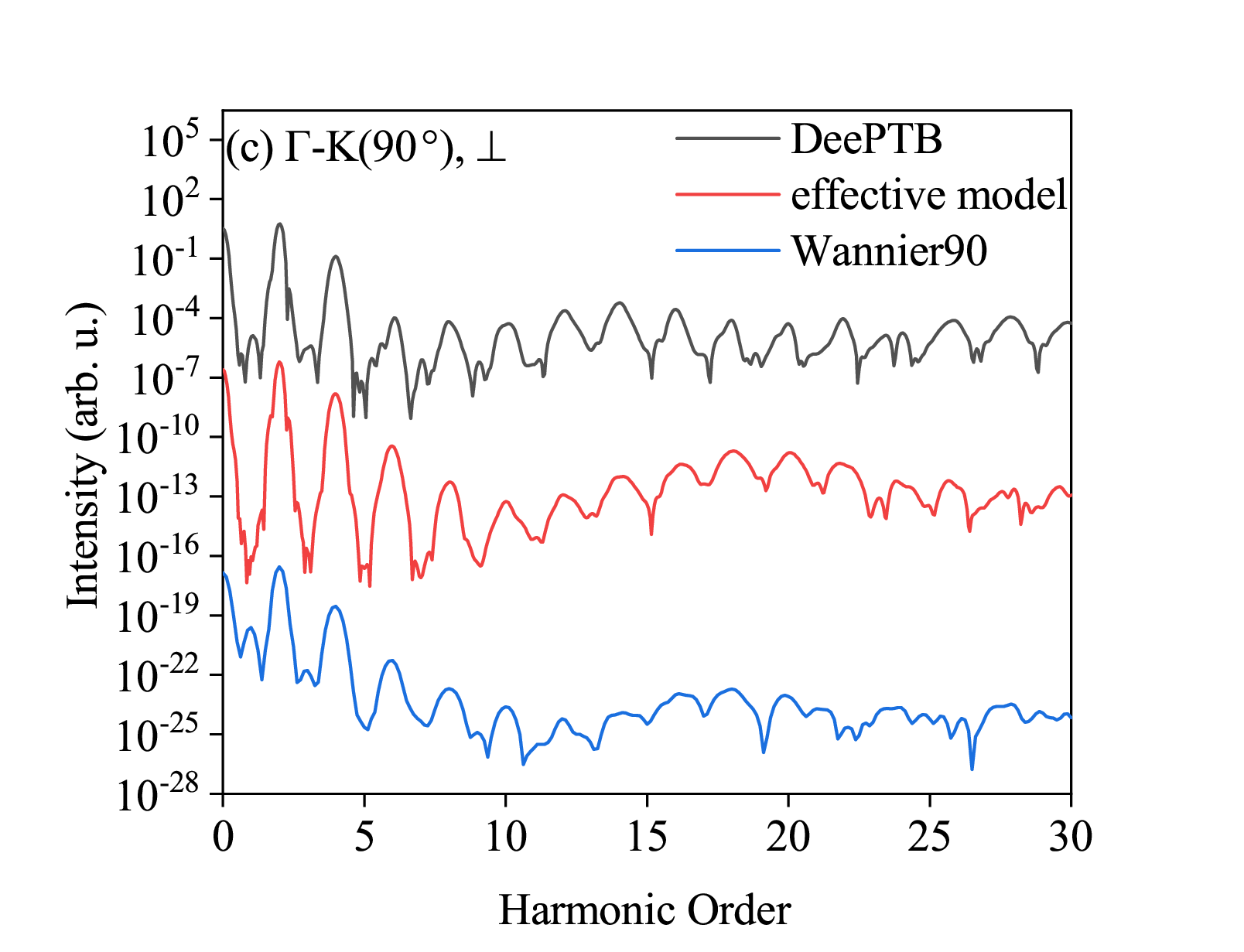}
}

\caption{
HHG spectra computed for a single monolayer of h-BN by the DeePTB (grey lines), the effective model (red lines) \cite{rn28}, and the Wannier90 based Hamiltonian, respectively. In (a) the laser pulse and the harmonic signal are polarized along the $\Gamma$-M direction. In (b) and (c), the laser pulse is polarized along the $\Gamma$-K direction, and the harmonic signals parallel and perpendicular to the this direction are presented, respectively. For clarity, the HHG spectra are vertically offset. 
}
\label{fig:hbn_hhg}
\end{figure*}

Using the trained Wannier-gauge DeePTB Hamiltonian, the HHG responses along different crystallographic directions were further calculated and compared with previous theoretical results obtained from other approaches. The simulations were performed using a laser pulse with a peak field strength of \(\mathrm{E}_0=\)40 MV/cm, a wavelength of 3 \(\mu{m}\), and a full width at half maximum (FWHM) of 34.2 fs. A \(100 \times 100 \times 1\) Monkhorst-Pack \(k\)-point grid was adopted, and the \(k\)-point sampling was verified to be converged. The pulse envelope was described by a \(\cos^2\) function, and the dephasing time is 2 fs. To obtain clearer harmonic spectra, a Hanning window was applied to the calculated current before Fourier transformation. We use the method described in Sec. \ref{subsection2B} to calculate the current and HHG of h-BN.

As shown in Fig. \ref{fig:hbn_hhg}, the HHG spectra along different directions driven by a linearly polarized laser are calculated by using the DeePTB-predicted Hamiltonian. They are in excellent agreement with the results obtained using the methods in Ref. \cite{rn28}. The effective model \cite{rn28} means the widely used simple tight-binding model in which only $p_z$ orbitals are considered. The Wannier90 refers to the results in which the Hamiltonian for calculating the HHG was obtained directly by the Wannier90 software \cite{MOSTOFI2008685}. The DeePTB method reproduces both the main harmonic peak positions and overall spectral characteristics. This agreement demonstrates the reliability of the machine-learning Hamiltonian for solid-state strong-field electron dynamics simulations and establishes its potential for studying more complex large-scale systems.

\subsection{Workflow for bulk Si calculations}
\begin{figure*}[!t]
\centering
\includegraphics[width=\textwidth,
    trim=10 150 90 15,
    clip]{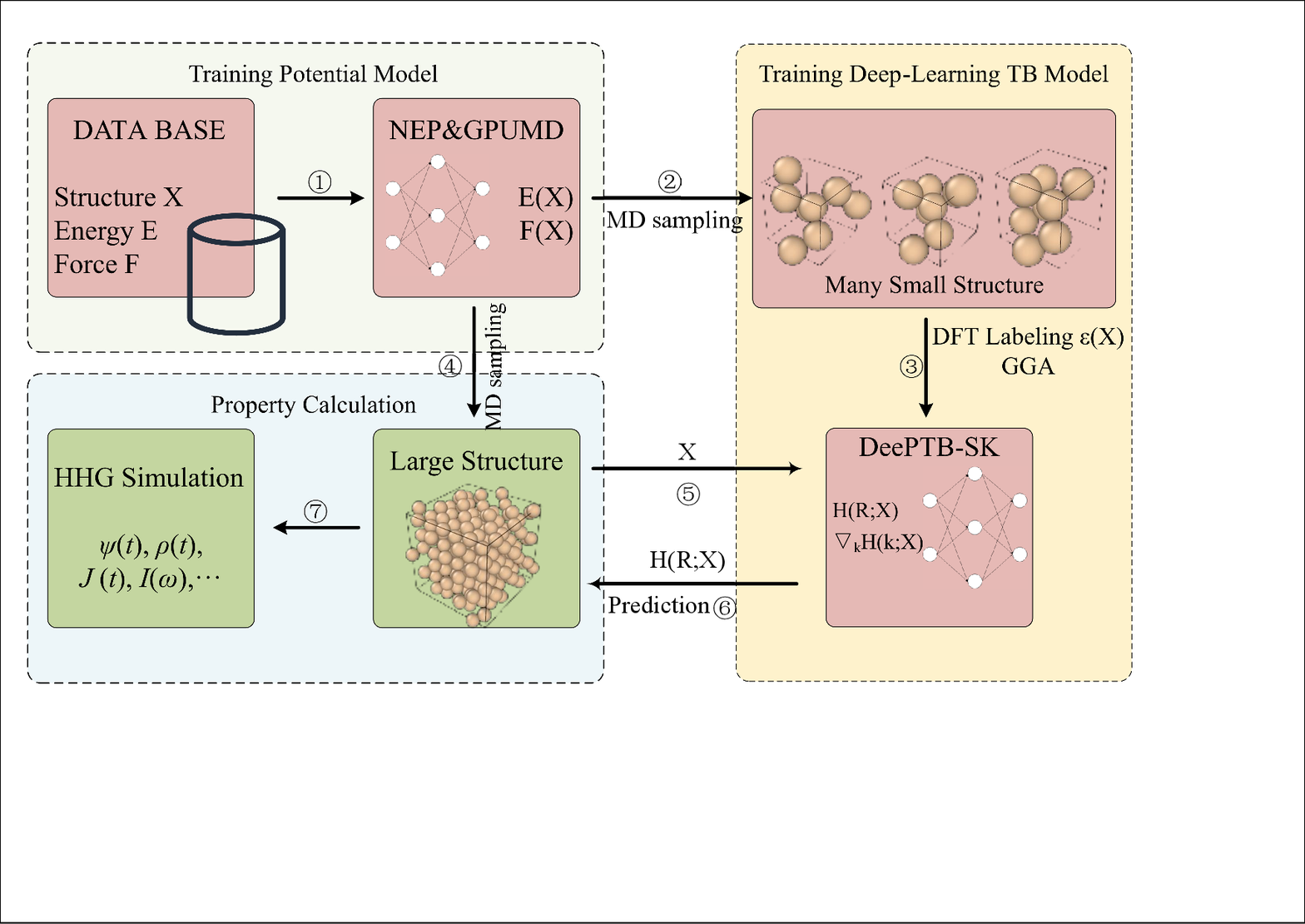}
\caption{\label{fig:workflow} Workflow for calculating phonon-assisted optical spectra using NEP, GPUMD and DeePTB models \cite{RU101}. NEP and GPUMD generate atomic structures through molecular dynamics simulations. Small-cell structures are then labeled by DFT eigenvalues to train the DeePTB model. Large-cell structures from GPUMD are then used with the trained DeePTB model to predict the Hamiltonians \(H(R)\). Based on
the Hamiltonians, current and HHG can be calculated. The NEP function we used is NEP89 \cite{rn40}. The steps in the flowchart are indicated by circled numbers.}
\end{figure*}

The workflow adopted in this study for the amorphous system is illustrated in Fig. \ref{fig:workflow}. Molecular dynamics simulations of Si unit-cell and supercell configurations at different temperatures are performed by using the GPUMD package \cite{rn30,rn31}. The simulations employed NEP89, a NEP model previously developed and trained in Refs. \cite{rn40,10.1063/5.0106617}.
High-throughput DFT calculations were subsequently carried out for the temperature sampled unit cell configurations. The DFT results from 50 temperature dependent configurations covering the range of 25\textendash 400 K were used as the dataset to train DeePTB \cite{RU101}. The trained model was then employed to predict the Hamiltonians of the supercell system at different temperatures. HHG calculations were performed based on these temperature-dependent Hamiltonians.

The details about our workflow are as follows. Initially, molecular dynamics simulations of bulk silicon unit cells are performed using GPUMD with NEP89 potential function over a temperature range of 25\textendash 400 K in canonical ensemble (NVT) and isothermal\textendash isobaric ensemble (NPT) to generate thermally distorted atomic configurations. The electronic band structures of these configurations are subsequently calculated using the ABACUS package. The resulting atomic configurations and their corresponding electronic band structures are then used to train the DeePTB model. Next, molecular dynamics simulations of 216-atom bulk silicon supercells are carried out using the same NEP89 potential and GPUMD over a temperature range of 100\textendash 300 K in NVT to sample thermally distorted supercell configurations at different temperatures. The trained DeePTB model is then employed to predict the corresponding tight-binding Hamiltonians for these configurations. Finally, the time-dependent electronic states are propagated using the Crank-Nicolson scheme, from which the time-dependent electric current and HHG are obtained.

\subsection{Electronic structure and strong-field response of bulk Si described by DeePTB}\label{subsection3C}

\begin{figure}[!tbp]
\includegraphics[width=\linewidth]{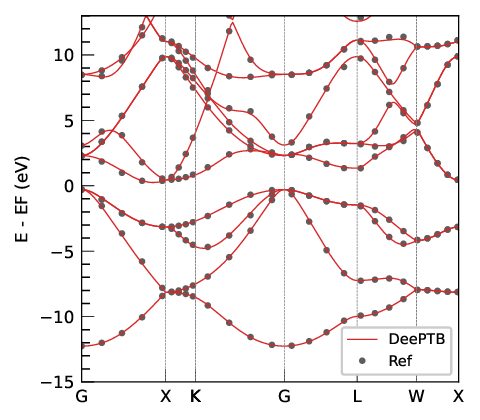}
\caption{\label{fig:band_si} Band structure of bulk Si
primitive cell by different methods. The red lines represent the band structure given by DeePTB, while the black dots correspond to the band structure calculated using ABACUS. EF means the Fermi energy. }
\end{figure}

Compared with the two-dimensional h-BN, bulk Si exhibits a more complex three-dimensional electronic structure and serves as an important system for studying strong-field responses in semiconductors. We performed DFT calculations using ABACUS with the pseudopotentials and orbitals from APNS to generate the electronic structure dataset of the bulk Si system \cite{rn39,rn40,rn41}, which was subsequently used to train the DeePTB model. We used the PBE functional for ABACUS calculations \cite{DD01}. First, the trained model described above was used to reproduce the band structure of the Si primary cell at 0 K, and the DeePTB results were compared with the DFT calculations by ABACUS \cite{rn39} as presented in Fig. \ref{fig:band_si}. The DeePTB Hamiltonian accurately reproduces the main electronic structure features of Si, demonstrating its capability for describing the electronic properties of the system.

\begin{figure}[!tbp]
\includegraphics[width=\linewidth,
    trim=20 20 40 20,
    clip]{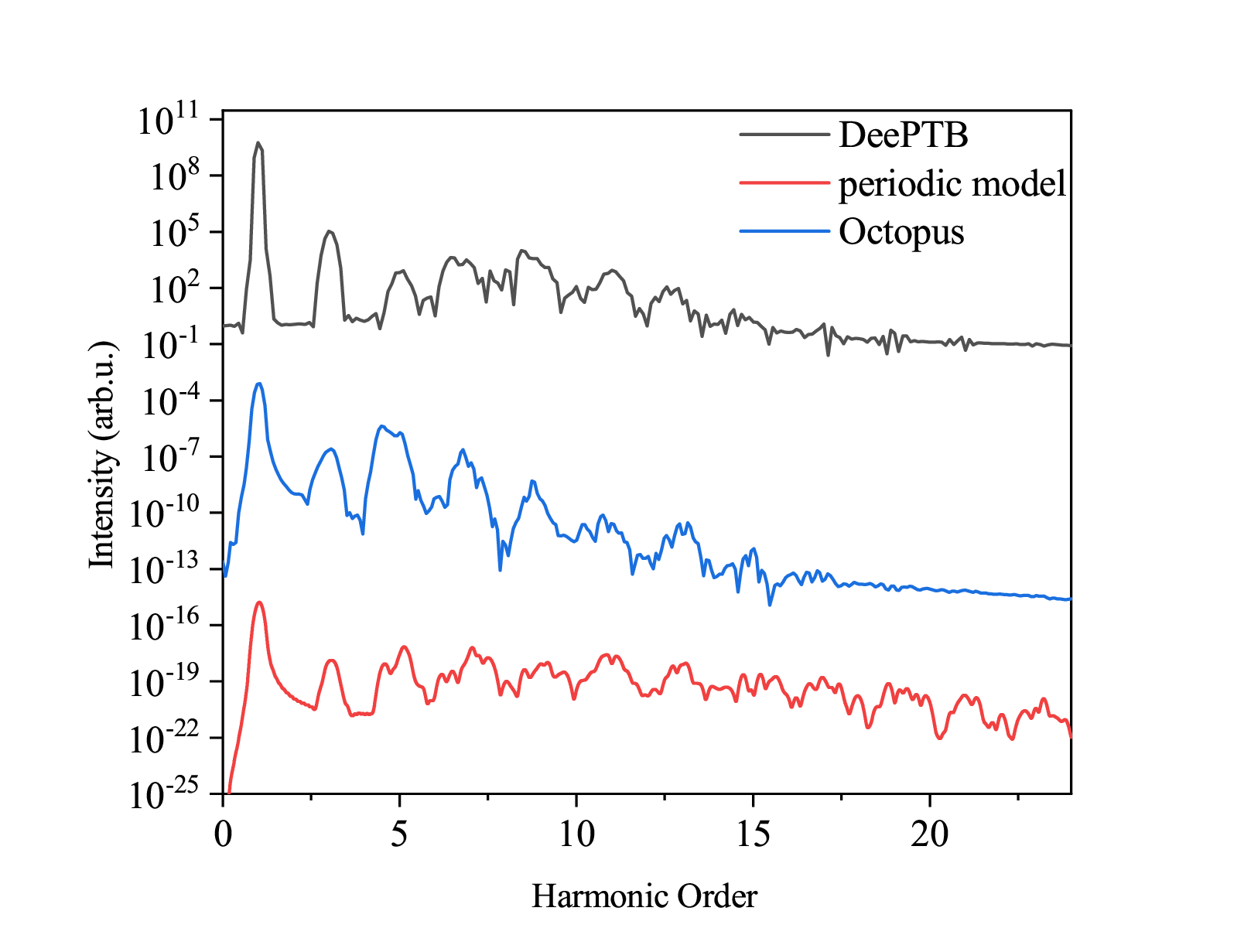}
\caption{\label{fig:Si_unit_cell_hhg} HHG of bulk Si unit cell. Grey line is HHG calculated based on DeePTB, red line is calculated by the periodic model \cite{10.1063/5.0314362}, and blue line is calculated by Octopus \cite{10.1063/5.0314362}. For clarity, the HHG spectra are vertically offset.}
\end{figure}

Subsequently, as shown in Fig. \ref{fig:Si_unit_cell_hhg}, the HHG response of the Si unit cell was calculated using the Hamiltonian by DeePTB and compared with previous TDDFT calculations by the software Octopus and periodic-model calculations in Ref. \cite{10.1063/5.0314362}. The real-space periodic model describes the motion of electrons in solids by introducing an effective periodic potential that reflects the periodicity of the crystal lattice. The simulations were performed using a laser pulse with a peak field strength of \(\mathrm{E}_0=10\,\mathrm{MV/cm}\), a wavelength of \(\lambda=2\,\mu\mathrm{m}\), and a FWHM of 30 fs. The vector potential of the laser field is given by

\begin{equation}
\begin{aligned}
\mathbf{A}(t)=
\frac{\mathbf{E}_0}{\omega}
\sin^2\left(\frac{\pi t}{2t_{\mathrm{FWHM}}}\right)
\sin(\omega t).
\end{aligned}
\label{eq:A(t))}
\end{equation}
The laser was incident normal to the crystal surface. A \(32\times32\times32\) Monkhorst-Pack \(k\)-point grid was adopted. In Subsections \ref{subsection3C} and \ref{subsection3D}, we use the method mentioned in Subsection \ref{subsection2C} to calculate the laser-induced current and HHG. The HHG spectra obtained using the DeePTB Hamiltonian are in good agreement with previous theoretical results \cite{10.1063/5.0314362}. Moreover, compared with the simple periodic model, the DeePTB results show closer agreement with the first-principles calculations.

Notably, this calculation requires 
approximately 8 hours on a single A100 GPU, demonstrating that the deep-learning tight-binding approach provides a favorable balance between accuracy and computational efficiency. This capability enables its application to large-scale systems in subsequent studies.
\subsection{Finite-temperature HHG response of large-scale Si supercells}\label{subsection3D}
\begin{figure*}[!tbp]
\centering

\subfloat{
    \includegraphics[
    width=0.32\textwidth,
    clip
    ]{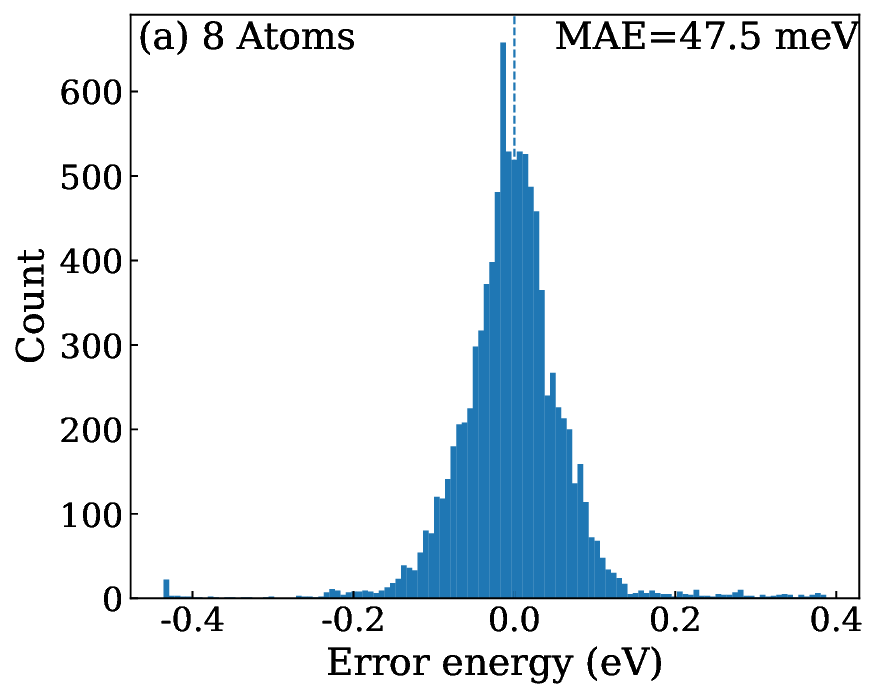}
}
\hfill
\subfloat{
    \includegraphics[
    width=0.32\textwidth,
    clip
    ]{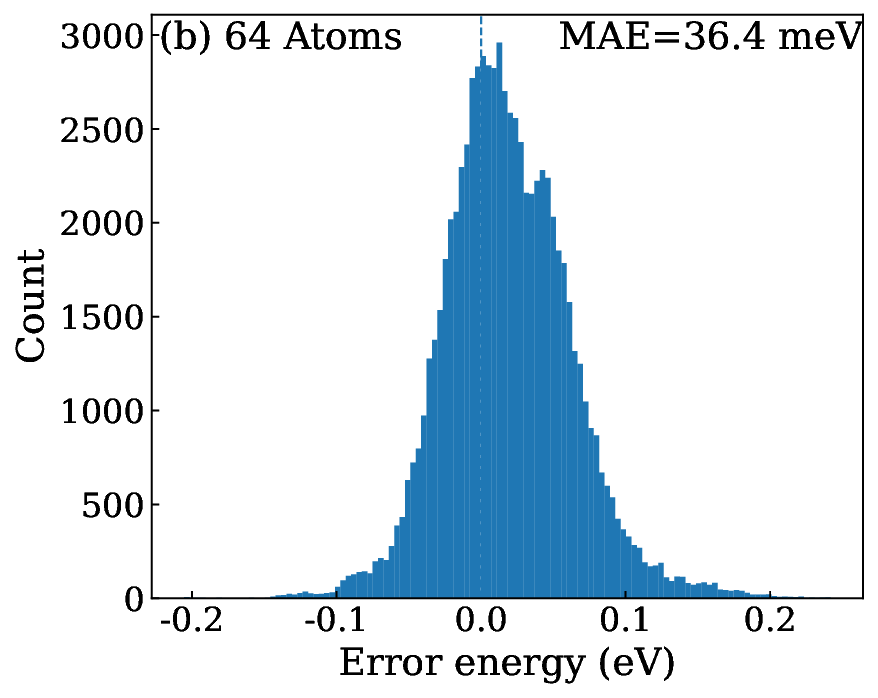}
}
\hfill
\subfloat{
    \includegraphics[
    width=0.32\textwidth,
    clip
    ]{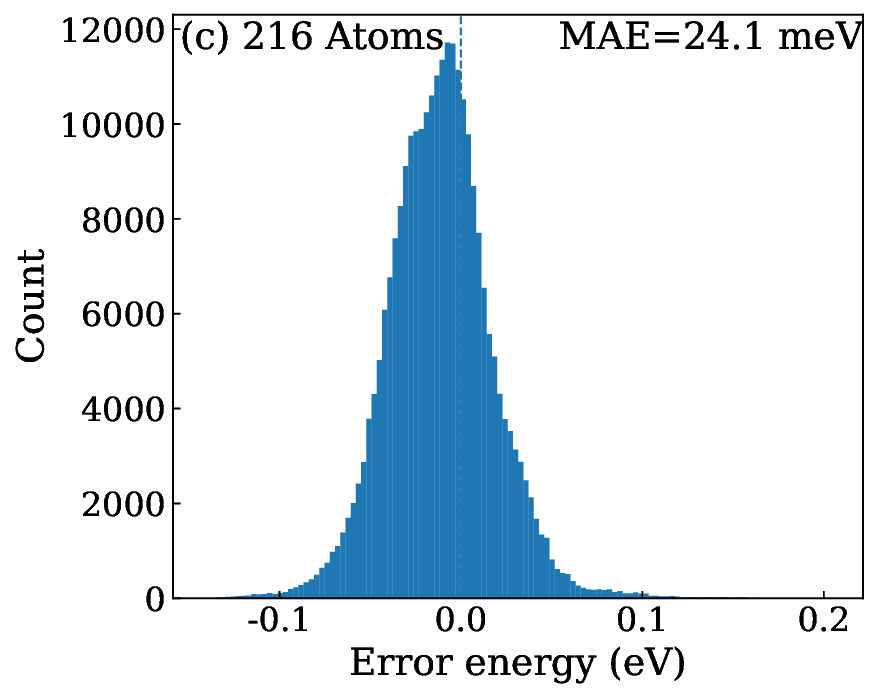}
}
\caption{\label{fig:band_error} Generalization and accuracy of DeePTB at different scales of bulk Si supercell. (a-c) The error histogram of bands between the results by DeePTB and the corresponding DFT calculations with a cell size of 8, 64, and 216 atoms, respectively. We count the energy differences for a number of bands equal to four times the number of atoms, sampling 302 $k$-points for each band.}
\end{figure*}

After establishing the reliability of the deep-learning tight-binding model for a unit cell, we further investigate the HHG response of large-scale Si systems under finite-temperature conditions. To assess the accuracy of DeePTB for Si supercells of different sizes, we compare its predicted band structures with those from DFT calculations at a fixed temperature of 300 K. The comparison is based on the distribution of the error energy and the mean absolute error (MAE). The error energy is defined as 
\begin{equation}
\Delta{E}=E_{{n}\boldsymbol{k}}^{DFT}-E_{{n}\boldsymbol{k}}^{DeePTB},
\label{eq:MAE}
\end{equation}
and MAE is defined as 
\begin{equation}
\textup{MAE}=\frac{1}{N_{\boldsymbol{k}}N_{band}}{\sum_{k=1}^{N_k}\sum_{n=1}^{N_{band}}|E_{{n}\boldsymbol{k}}^{DFT}-E_{{n}\boldsymbol{k}}^{DeePTB}|},
\label{eq:MAE}
\end{equation}
where \(N_{\boldsymbol{k}}\) is the number of $k$-points per band, \(N_{band}\) is the number of bands we mainly considered. \(E^{DeePTB}_{n\boldsymbol{k}}\) and \(E^{DFT}_{n\boldsymbol{k}}\) denote the energies of the $n$th band and $k$th $k$ point predicted by DeePTB and calculated by ABACUS, respectively. MAE can quantitatively demonstrate the predictive accuracy. Typically, MAE values lower than 50 meV indicate that the error is negligible compared with the typical energy scale of electronic bands \cite{1-20251253}. As shown in Fig. \ref{fig:band_error}, the MAE remains nearly unchanged when the system size increases, proving that DeePTB captures local atomic environment effects and can be generalized from small cells to large supercells. DeePTB maintains high predictive accuracy for Si supercells with various numbers of atoms, indicating that the method can effectively capture the variation of electronic structures induced by thermal fluctuations.

\begin{figure}[!tbp]
\includegraphics[width=\linewidth,
    trim=30 20 90 20,
    clip]{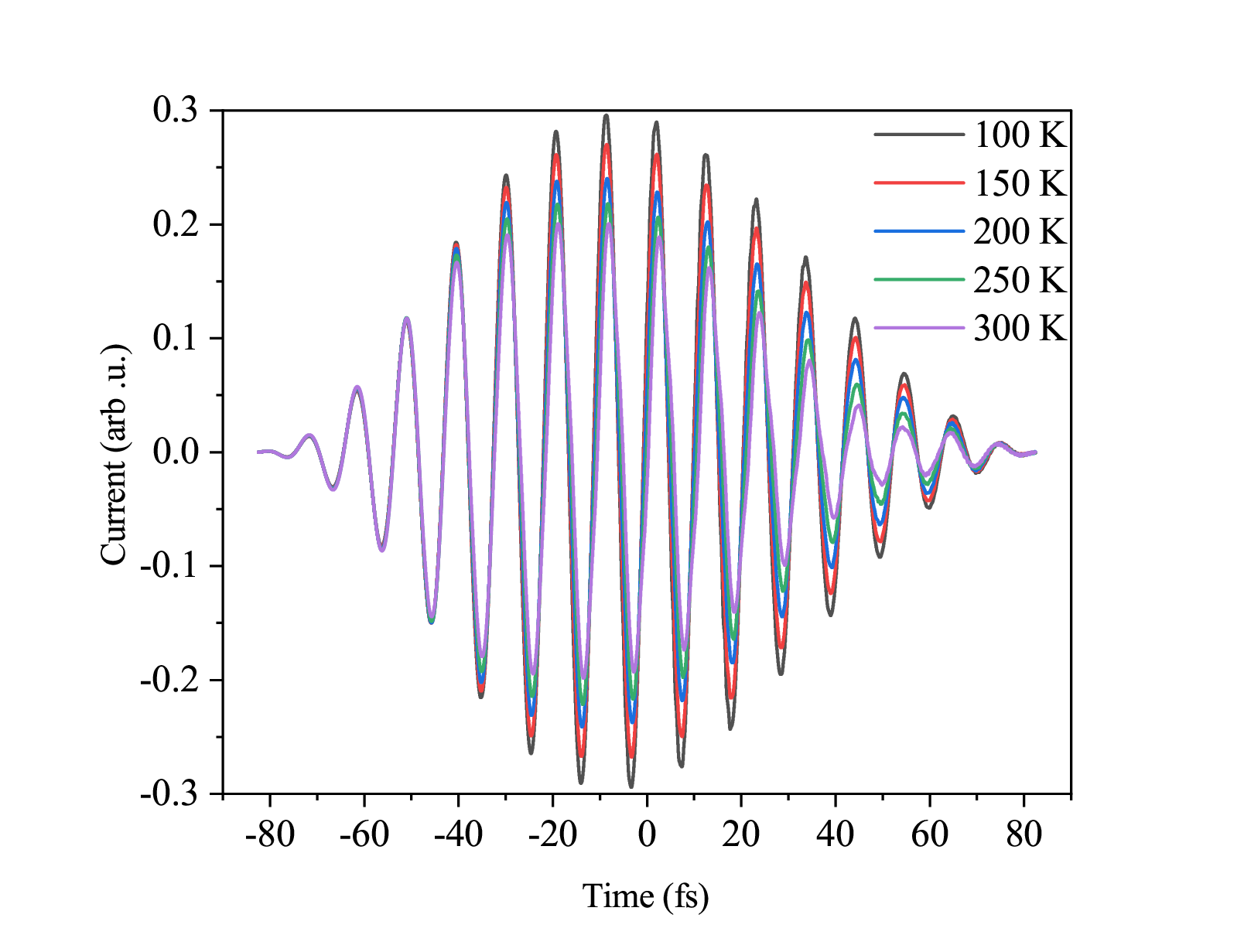}
\caption{\label{fig:Si_current} Laser-induced currents of a 216-atom bulk Si supercell at various temperatures. The gray, red, blue, green, and purple lines correspond to the current responses of the bulk Si supercell at 100 K, 150 K, 200 K, 250 K, and 300 K, respectively.}
\end{figure}

\begin{figure}[!tbp]
\includegraphics[width=\linewidth,
    trim=30 20 90 20,
    clip]{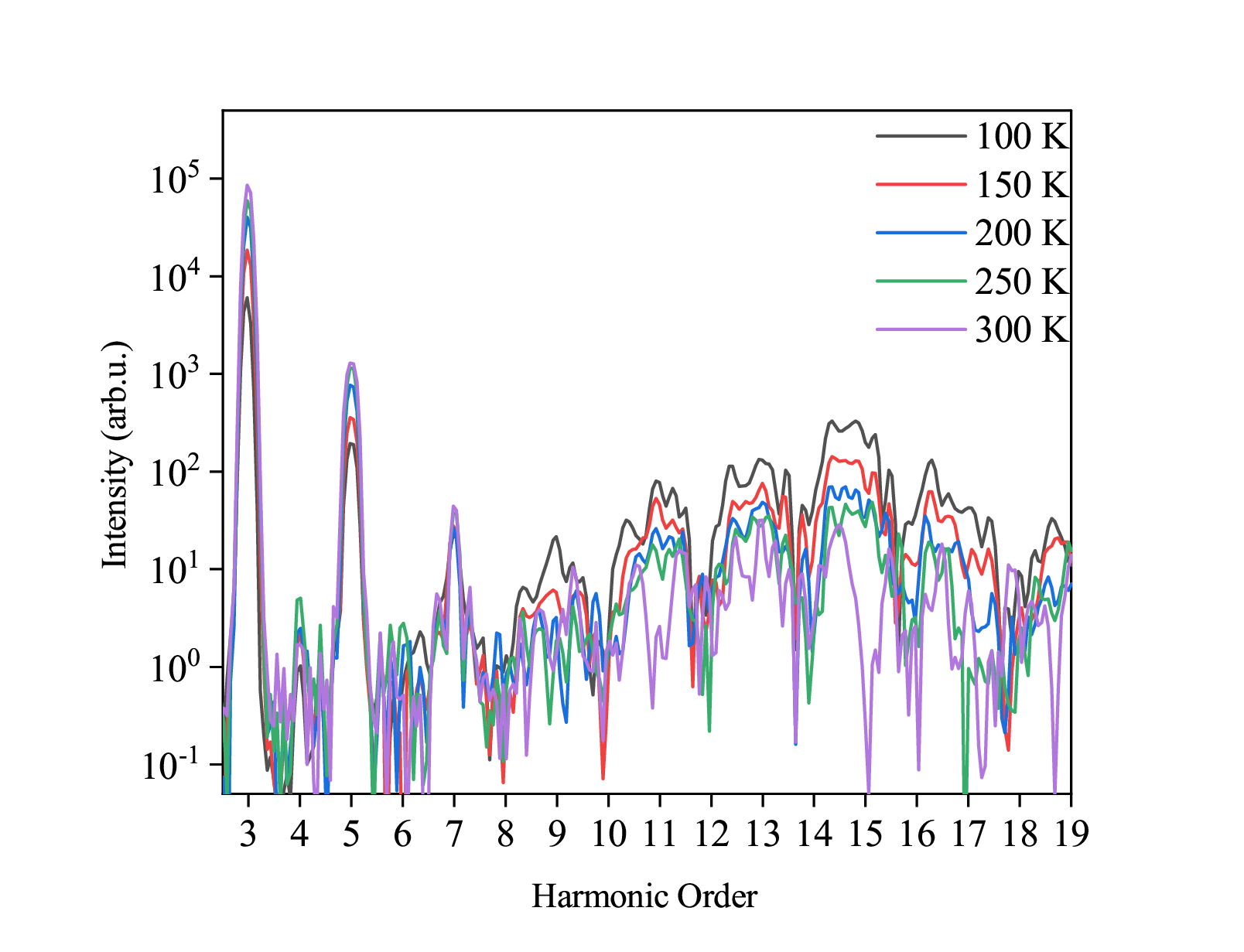}
\caption{\label{fig:Si_supercell_hhg} Calculated high harmonic spectra for different temperatures based on DeePTB.}
\end{figure}

Considering the balance between computational accuracy and efficiency, we select a Si supercell containing 216 atoms for the temperature-dependent HHG calculations. We employ laser pulses with a peak electric field strength of \(\mathrm{E}_0=15\,\mathrm{MV/cm}\), a wavelength of \(\lambda=3.2\,\mu\mathrm{m}\), an FWHM of \(60\,\mathrm{fs}\), and a \(\cos^2\) pulse envelope. The laser pulse is incident perpendicular to the crystal surface. The time-dependent current responses at different temperatures are calculated using the SBE based on occupied-state propagation. A
 \(11\times11\times11\) Monkhorst-Pack $k$-point grid was adopted, and five atomic configurations are sampled for each temperature. The results have been verified to achieve convergence. And this calculation requires approximately
4 days on a single A100 GPU for one supercell structure.
 
Figure \ref{fig:Si_current} presents the time-dependent current responses of Si supercells at different temperatures. With increasing temperature, the electronic dynamics driven by the strong laser field exhibit pronounced changes. The amplitude of the current decreases as the temperature increases, demonstrating that thermal structural fluctuations can effectively modulate the strong-field electron dynamics. We further perform a Fourier transform of the time-dependent current to obtain the HHG spectra at different temperatures. A Hanning window is applied to the current signal to improve the clarity of the harmonic spectra. As shown in Fig. \ref{fig:Si_supercell_hhg}, low-order and high-order harmonics exhibit distinct temperature-dependent behaviors. For the third- and fifth-order harmonics, the harmonic yield gradually increases with increasing temperature. In contrast, the harmonic yield of higher-order harmonics (ninth order and above) decreases as the temperature increases. The temperature-dependent behavior of different harmonic orders exhibits a transition around the seventh harmonic order. This is in good qualitative agreement with the experimental measurements of HHG in Si for the crystalline and amorphized states \cite{2604.07543, purschke2025giantenhancementattosecondtunnel}, further indicating the validity of our method.

\begin{figure}[!tbp]
\includegraphics[width=\linewidth,
    trim=20 50 90 20,
    clip]{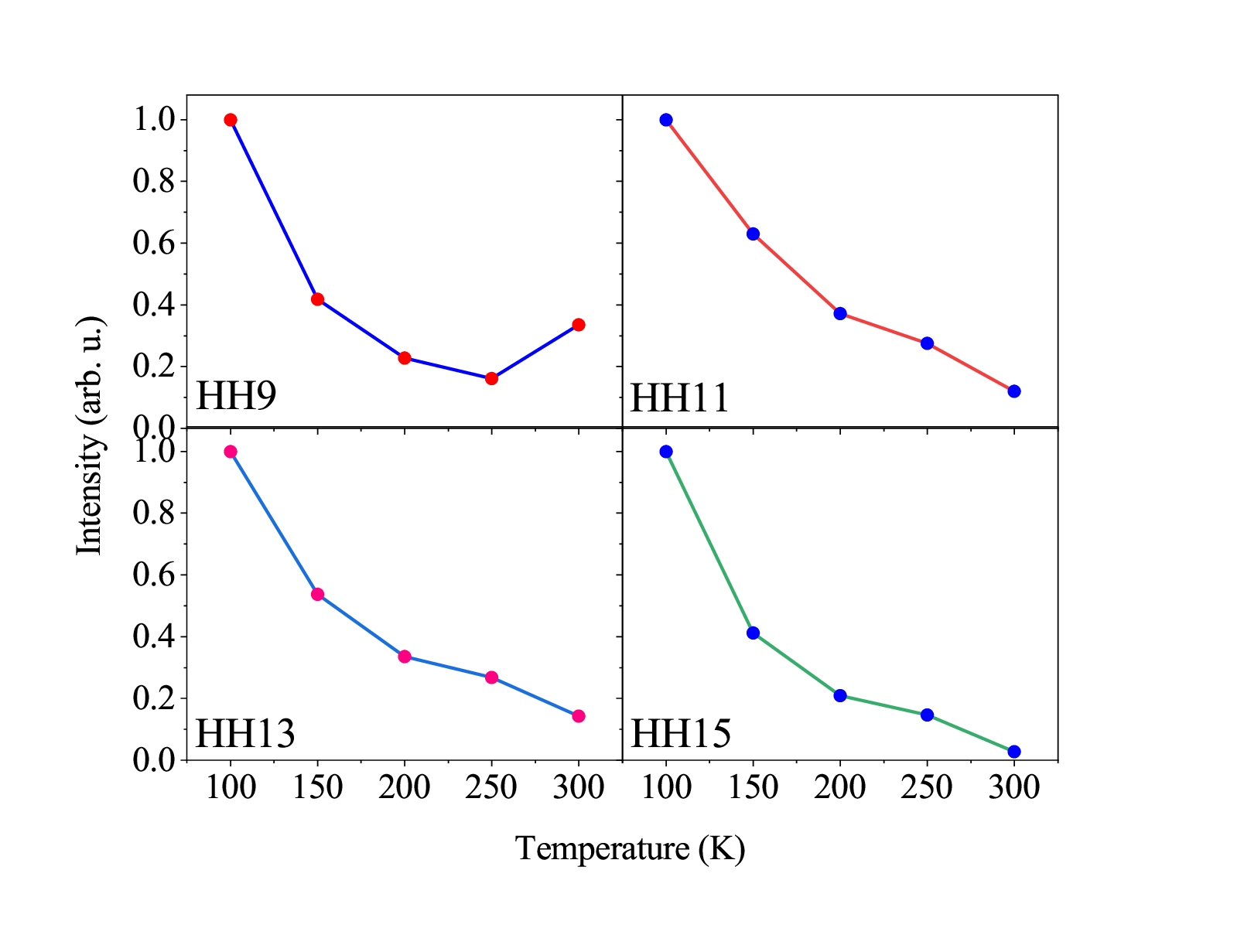}
\caption{\label{fig:Si_high_order} 
Temperature dependence of the harmonic yield for the 9th (a), 11th (b), 13th (c), and 15th (d) harmonics. The points are simulation results at 100 K, 150 K, 200 K, 250 K, and 300 K (Use 100 K as the baseline value of 1), respectively.}
\end{figure}

\begin{figure}[!tbp]
\includegraphics[width=\linewidth,
    trim=0 165 0 0,
    clip]{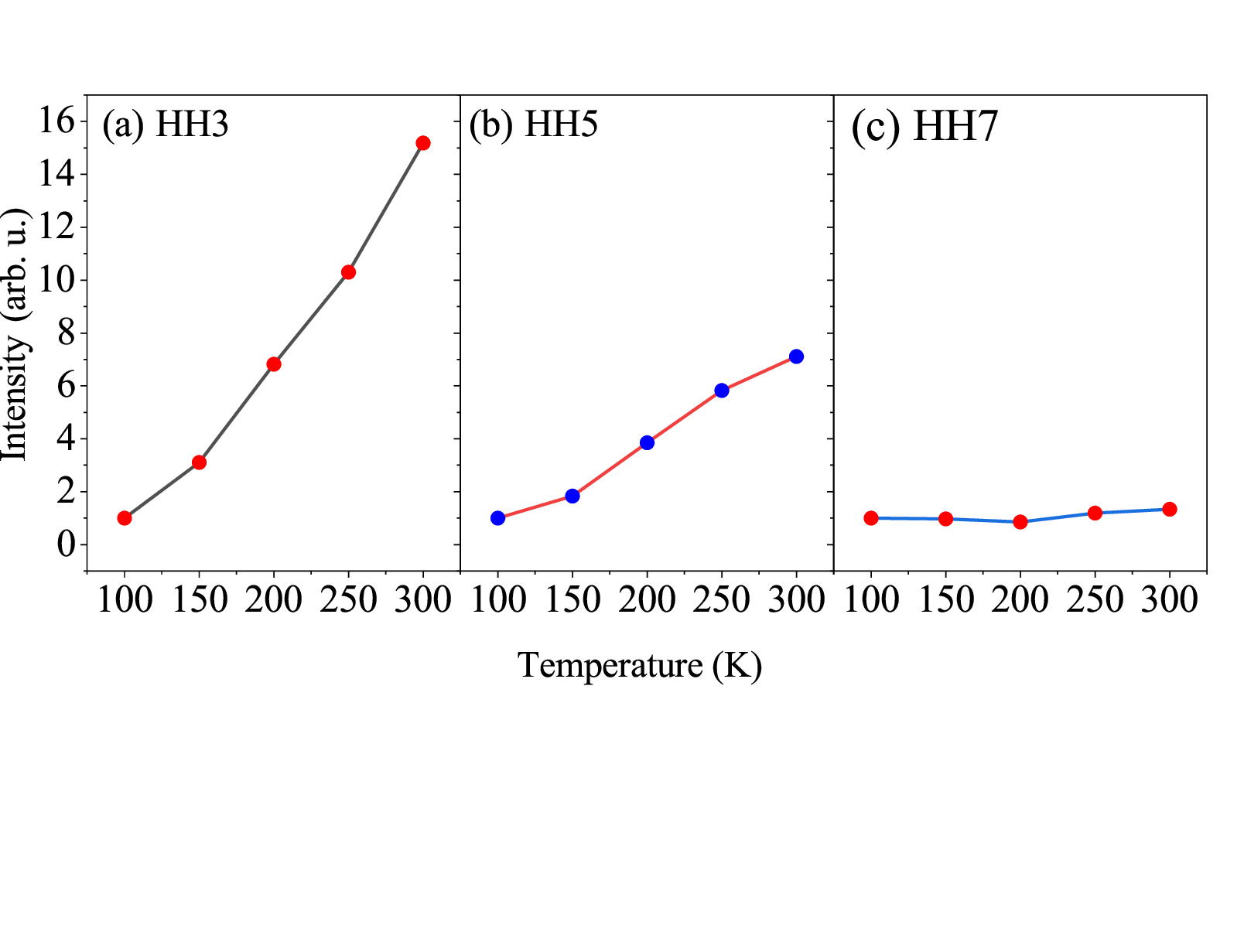}
\caption{\label{fig:low_order} Temperature dependence of the harmonic yield for the 3rd (a), 5th (b), 7th (c) harmonics. The points are simulation result at 100 K, 150 K, 200 K, 250 K, and 300 K (Use 100 K as the baseline value of 1), respectively.}
\end{figure}

To quantitatively analyze the temperature dependence of different harmonic orders, we calculate the variation of the harmonic yield for each order as a function of temperature. To further illustrate this trend, Figs. \ref{fig:Si_high_order} and \ref{fig:low_order} show the normalized temperature dependence of the high-order and low-order harmonic yields, respectively, with respect to the results obtained at 100 K. 

\begin{figure}[!tbp]
\includegraphics[width=\linewidth,
    trim=40 20 90 20,
    clip]{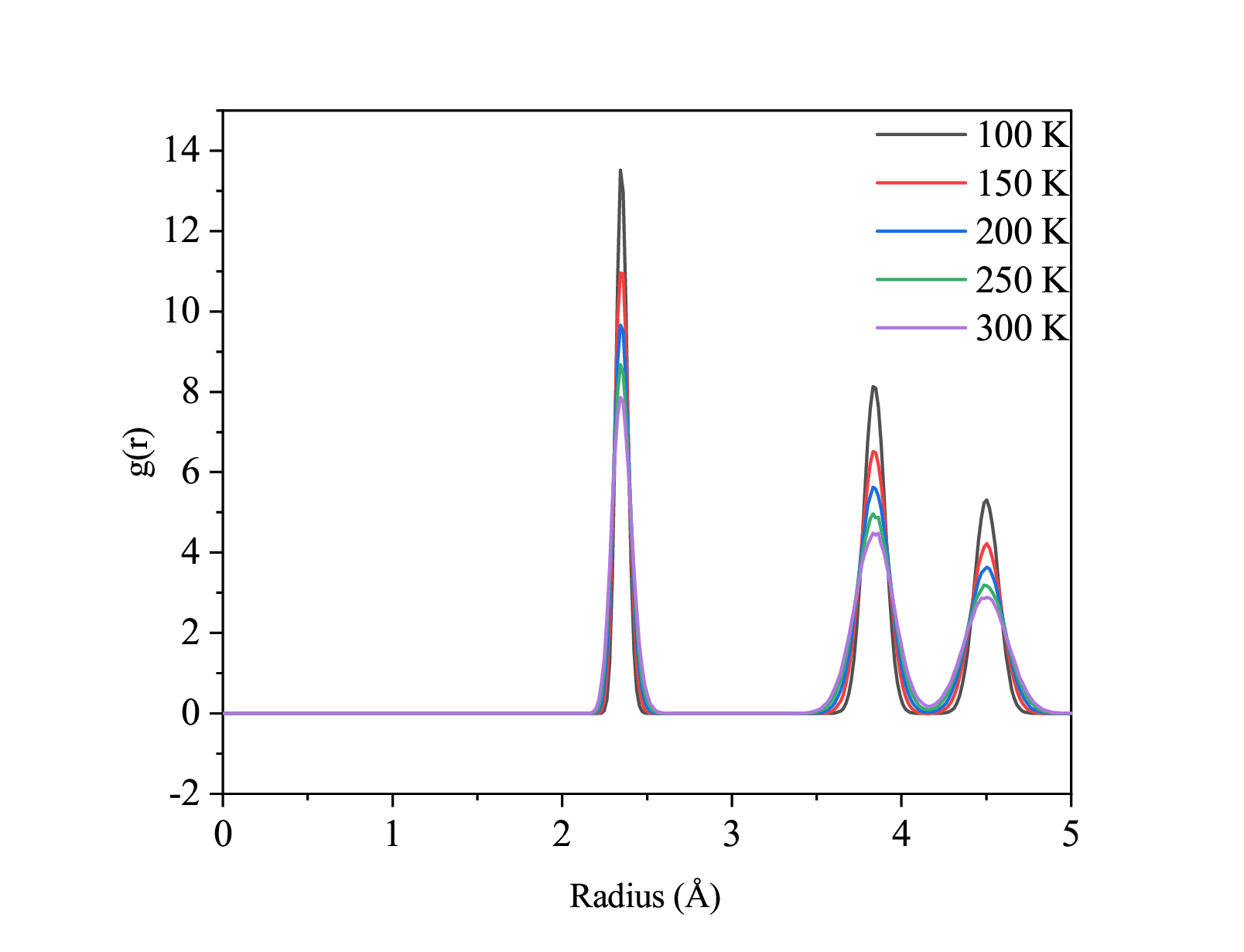}
\caption{\label{fig:rdf} The radial distribution functions of bulk Si at 100 K (grey line), 150 K (red line), 200 K (blue line), 250 K (green line), 300 K (purple line), respectively.}
\end{figure}

We discuss the high order harmonics first. The results presented in Fig. \ref{fig:Si_high_order}  are in good agreement with the previously reported experimental results \cite{2604.07543}.
As the temperature increases, thermal atomic displacements become more pronounced, leading to enhanced structural fluctuations in the Si supercell. The radial distribution functions shown in Fig. \ref{fig:rdf} reveal the temperature-dependent structural variations of bulk Si. With increasing temperature, the peaks of the radial distribution functions become broader and less pronounced, indicating a wider distribution of Si\textendash Si bond lengths caused by enhanced atomic vibrations. Meanwhile, the attenuation of higher-order peaks demonstrates a reduction of long-range structural correlations due to thermal disorder. These features directly evidence that elevated temperatures introduce stronger lattice fluctuations in the Si supercell. These thermal structural fluctuations modify the electronic Hamiltonian and consequently influence the strong-field-driven electron dynamics.

\begin{figure*}[!tbp]
\centering
\includegraphics[width=\linewidth,
    trim=0 190 0 0,
    clip]{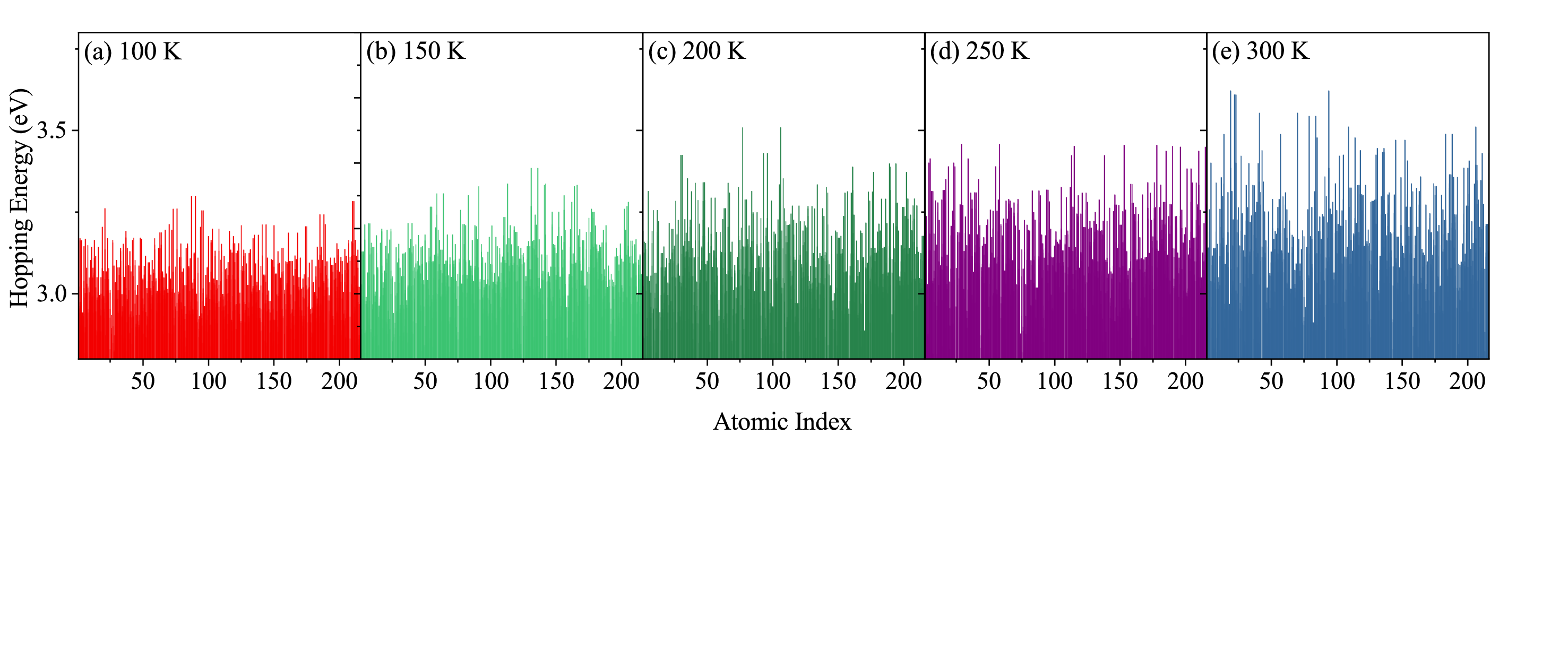}
\caption{\label{fig:hoping energy} Histograms of interatomic orbital hopping energy distributions in a 216 atoms bulk Si supercell at 100 K(a), 150 K(b), 200 K(c), 250 K(d) and 300 K(e), respectively. The atomic index labels all Si atoms in the supercell, and the displayed hopping energy is the maximum hopping energy associated with each atom as the central site.}
\end{figure*}

Because bulk Si has a small thermal expansion coefficient, lattice expansion can be safely neglected within the temperature range of 100\textendash 400 K under ambient pressure \cite{rn46,RN47,RN48}. Therefore, the temperature dependence of harmonic yields mainly originates from thermally induced structural fluctuations rather than lattice expansion effects \cite{RU101}. Because lattice expansion is negligible in this regime, the variation in HHG yields can be attributed mainly to temperature-induced atomic rearrangement. In finite-temperature solids, lattice vibrations influence electronic dynamics through electron-phonon coupling. In this work, instead of introducing an additional empirical electron-phonon scattering model, we explicitly generate thermally perturbed crystal structures through finite-temperature molecular dynamics simulations and construct temperature-dependent electronic Hamiltonians based on these structures. The high-order harmonics are mainly generated from long-range electron transport. These thermally perturbed structural disorder induces scattering which changes the coherence of electron-hole pairs during their excursion in laser-induced dynamics \cite{2604.07543}. Thus the higher-order harmonics become weak as temperature increases. In addition, we discuss the low-order harmonics. They are in the perturbative regime. The excursion of corresponding electron-hole pairs may be negligible. To gain deeper insight into the microscopic origin of the temperature-dependent enhancement of low-order harmonics, we further investigate the evolution of interatomic hopping amplitudes within the Si supercell. As illustrated in Fig. \ref{fig:hoping energy}, the hopping amplitudes exhibit an overall enhancement with increasing temperature. This behavior is also closely related to the thermally induced structural fluctuations revealed by Fig. \ref{fig:rdf}. As discussed above, the peaks of the radial distribution functions become progressively broad with increasing temperature, while the characteristic Si\textendash Si bond lengths remain nearly unchanged. This observation suggests that thermal excitation primarily increases the fluctuations of local atomic configurations rather than producing a substantial change in the average crystal structure. Such fluctuations continuously modify the spatial separation and relative orientation of neighboring atomic orbitals, thereby altering their overlap and the corresponding electronic hopping amplitudes. Consequently, even in the absence of appreciable lattice expansion or contraction, enhanced thermal vibrations can produce noticeable modifications of the local electronic coupling within the supercell.

The microscopic origin of this behavior can be further understood from the nonlinear dependence of interatomic hopping on bond length \cite{EE01}. Within the tight-binding framework, hopping amplitudes describe the coupling between neighboring atomic orbitals and are highly sensitive to their spatial overlap. In general, decreasing the interatomic distance increases the orbital overlap and tends to enhance the magnitude of the corresponding hopping integral, whereas increasing the distance produces the opposite effect. Importantly, these two contributions are not necessarily symmetric because the hopping amplitude depends nonlinearly on the interatomic separation \cite{HH01,HH02}. Because hopping magnitudes respond nonlinearly to the changes in bond length, the enhancement associated with shortened bonds can outweigh the reduction caused by elongated bonds. As temperature increases, the broadening of the Si\textendash Si bond-length distribution increases the occurrence of both compressed and elongated local configurations. The resulting asymmetric response of the hopping amplitudes may therefore lead to an increase in their statistical average, even when the characteristic bond length remains nearly constant. In addition, thermally induced variations in local bond orientations can modify the directional overlap and hybridization of Si orbitals, further contributing to the redistribution of electronic coupling strengths. These effects provide a plausible microscopic explanation for the observed temperature-dependent enhancement of interatomic hopping amplitudes.

Importantly, the hopping amplitudes determine the coupling between atomic orbitals and contribute to the electronic dispersion and the associated velocity and current operators. Consequently, temperature-induced modifications of hopping amplitudes can affect the response of electrons to an external laser field \cite{RN2020}. In the strong-field regime, changes in local electronic coupling can affect carrier dynamics, electron\textendash hole coherence, and related quantities. Such modifications can alter the efficiency of harmonic emission. Therefore, the observed enhancement of low-order harmonic yields may be associated with the nonlinear modulation of interatomic electronic hopping by lattice thermal fluctuations. This interpretation suggests that the temperature dependence of low-order HHG is sensitive not only to changes in electronic energy levels but also to the microscopic redistribution of orbital coupling induced by thermal atomic motion.

In contrast, high-order harmonics rely on phase-coherent electron motion over extended spatial regions under strong-field excitation. Although thermal structural fluctuations enhance local orbital coupling, they also introduce temperature-dependent variations into the electronic Hamiltonian and perturb the phase coherence of long-range electron dynamics. This dephasing effect induced by the structural disorder reduces the efficiency of high-order harmonic, resulting in a temperature dependence opposite to that of low-order harmonics.

These results demonstrate that lattice thermal fluctuations at finite temperatures do not simply enhance or suppress HHG. Instead, they induce a harmonic-order-dependent redistribution of the HHG response by simultaneously modifying local electronic coupling and strong-field-driven coherent electron dynamics. The enhanced local coupling promotes low-order harmonic emission, whereas the modification of phase-coherent electron motion reduces the efficiency of high-order harmonics, leading to distinct temperature responses across different harmonic orders.
\section{SUMMARY}\label{section4}
Lattice thermal fluctuations are ubiquitous in solids, yet their role in strong-field HHG remains difficult to investigate in realistic large-scale systems. Here we develop an approach that combines deep learning molecular dynamics, the DeePTB electronic Hamiltonian, and semiconductor Bloch equations. Calculations for h-BN show that the DeePTB Hamiltonian reproduces the reference HHG spectra with high accuracy. We then apply the method to 216-atom bulk Si supercells using an occupied-state propagation scheme to study HHG in thermally fluctuating lattices.

We find that lattice temperature modifies the harmonic response in an order-selective manner. The intensities of low-order harmonics increase with temperature, whereas those of high-order harmonics decrease, with a crossover near the seventh harmonic. By analyzing the temperature-dependent electronic Hamiltonians, we trace the enhancement of low-order harmonics to thermally induced changes in local orbital couplings and the resulting increase in hopping amplitudes. For the higher harmonic orders, thermal structural fluctuations instead suppress the coherence of the strong-field electron dynamics in a different regime, leading to a contrasting temperature dependence.

These results show that lattice thermal fluctuations can selectively reshape solid-state HHG rather than simply suppressing the harmonic response. The combination of machine-learning electronic Hamiltonians and SBE provides an efficient route for studying temperature-dependent strong-field electron dynamics in large and structurally disordered solid systems.

\begin{acknowledgments}
We thank Prof. S. Gräfe and Dr. M. Thümmler for providing the data obtained with Octopus and the periodic model. We thank Prof. T.T. Luu, Dr. J.C. Zhang, Prof. M.H. Chen, Prof. Q.Q. Gu, Dr. Z. Zhouyin, Prof. Z.Y. Fan, Prof. J.X. Chen, and Dr. Z.W. Ding for helpful discussions. This work is supported by the Hubei Provincial Natural Science Foundation of China (Grant No. 2024AFA029), the National Natural Science Foundation of China (Grant Nos. 12304383, and 12674429), and the Hubei Province Postdoctoral Pioneer Support Program (Grant No. 2025HBBSHXF011).

\end{acknowledgments}
\bibliography{Si} 

\end{document}